\documentstyle{elsart} 
 
\begin{document} 
 
\input{epsf.tex} 
 
\begin{frontmatter} 
 
\title{TOTAL AND NUCLEAR PHOTOABRORPTION CROSS SECTIONS OF $^{52}Cr$ 
\vskip 0 cm 
IN THE ENERGY RANGE OF 8-70 MEV }

\author[inr,knu]{V.Kuznetsov\thanksref{slava}}, 
\author[inr]{S.Merkulov}, 
\author[inr]{G.Solodukhov\thanksref{gv}}, 
\author[inr]{Yu.Sorokin}, 
%\author[inr]{B.Tulupov}, 
\author[inr,kur]{A.Turinge}.

\address[inr]{Institute for Nuclear Research, 117312 Moscow, Russia} 
\address[knu]{Kyungpook National University, 702-701, Daegu, Republic of Korea}
\address[kur]{RRC "Kurchatov Institute", Moscow, Russia}

\thanks[slava]{Contact person. E-mail: slava@cpc.inr.ac.ru, slavaK@jlab.org} 
\thanks[gv]{Contact person. E-mail: solod@inr.ac.ru} 
 
\date{\today}

%-------------abstract---------------- 
 
\begin{abstract} 
 
Total (atomic+nuclear) photoabsorption cross section of $^{52}Cr$  
was for the first time measured in the energy range of 8-70 MeV. 
Experimental data was produced with small statistic 
and systematic errors.  The results deviate from calculations 
of atomic cross section at photon energies above 40 MeV. 
Photonuclear cross section in the region of the E1 giant dipole resonance 
(GDR) clearly exhibits three peaks at 18.9, 20.9 and 23.1 MeV. 
At higher energies, the measured cross section hints a dip-peak structure 
at 40-48 MeV. 
 
\end{abstract} 
 
%----------end of abstract------------- 
 
\end{frontmatter}

%-------------main text---------------- 

\section{Introduction} 
 
Detailed study of photonuclear interaction in the region of E1 and E2 
resonances  provides meaningful information regarding  
fundamental properties of nuclear matter. Since electromagnetic 
interaction is well known, the interpretation of such experiments is less complicate in respect to 
hadron- and meson-nuclear interaction. 
 Several problems are of special interest in this physics domain:   
i) the evolution of the giant dipole E1 resonance (GDR) properties with the mass number $A$; 
ii) the relative role of the isospin and congigurational GDR splittings; 
iii) contributions of resonances others than E1, in particular, 
the isoscalar E2 resonance. 

Given these points, the study of 
medium nuclei with mass number $50 \leq A \leq 100$ is of particular interest.
The GDR in these nuclei is concentrated in a rather  
narrow energy range  while the expected position of the  
$E2,T=0$ resonance is on the growing part of the E1 dipole  
cross section. Both factors favor the discrimination of one resonance from the 
other in this mass region. In heavy nuclei,   
the E2 isoscalar resonance is located just near the $(\gamma, n)$
threshold (6-8 MeV), and its study is more difficult
through the small cross section.  For light nuclei, the complicated fine 
structure of the GDR requires high resolution detectors, and, in addition, 
the interpretation of many overlapping peaks in photonuclear cross section  
is rather sophisticated.  
 
Similar consideration can be applied to the study of the isospin splitting of 
giant resonances.  Observation of this effect is more difficult 
in case of heavy nuclei since  $\Delta T=1$ decays of resonances 
are suppressed in the photoproton  channel by Coulomb forces, while 
in the photoneutron channel they are forbidden by the isospin conservation. 
 
The experimental investigation of the configurational splitting of the GDR is the most important task 
for the understanding of the GDR excitation. The GDR configurational 
splittings was first predicted  for light nuclei in Ref.~\cite{neu} and 
then proved in Ref.~\cite{era}. Up to now it was observed only in 
the region of nuclei with   with $16<A<40$~\cite{gab,gsi}. An open question
is whether it can play a role at higher A.  

Reliable experimental data for medium $50 \leq A \leq 100$ nuclei  
are essential to determine the limit of validity of theoretical models 
describing nuclear excitations. Microscopic calculations  
are difficult for such many-nucleon systems, while collective models, 
well describing heavy nuclei, have not been yet extended to medium
region because of lack of experimental data.  
 
Measurements of total photonuclear cross section ($\sigma _{tot_{nucl}}$)
is quite attractive for the interpretation of experimental results since 
it doesn't require any assumptions on the nuclear de-excitation. 
Therefore the experimental data on $\sigma _{tot_{nucl}}$ for  medium nuclei may 
provide a challenge for the further development  of theoretical models. 
At present the available data base for total and nuclear photoabsorption 
cross sections comprises only data points for light and heavy nuclei. Light nuclei  
($A \leq 30$) have been intensively studied in pioneer experiment
\cite{luba1} that aimed to establish a fine structure of the GDR. 
Later, experimental efforts have been extended to the heavy-nuclei region 
($A\sim 200$) \cite{sol5,sol6}. The goal of those measurements was to 
observe the evolution of GDR parameters with the mass number A. 
There is a lack of data for medium nuclei.  
    
\section{Attenuation method} 
 
There are two main methods to obtain $\sigma _{tot_{nucl}}$: \\ 
{\it 1. Summing of cross sections of main partial channels}. This method was 
widely used in studies of heavy nuclei. For these nuclei 
the emission of a single neutron is a dominating channel. However, 
for light nuclei the contribution of the $(\gamma, p)$ reaction  
becomes more essential, and, for some nuclei, exceeds the $(\gamma, n)$ one. 
The efficient detection of the $(\gamma, p)$ final state  
in not easy in experiment because of the internal absorption of emitted  
protons inside the target. \\ 
{\it 2. Attenuation method}. This method implies on a measurement 
of the attenuation of a photon beam by a sample of material under study. 
Attenuation coefficients at different photon energies are obtained as  
a ratio of two spectra:  a spectrum of a beam attenuated by a sample of material
and a spectrum of an unattenuated beam. Total (atomic-plus-nuclear) 
photoabsorption cross section is derived from the attenuation coefficients.
Further photonuclear cross section can be derived as a difference between  
the measured total cross section and the calculated atomic one.   
This method, being more accurate, requires some specific experimental 
efforts: a high-resolution and stable photon spectrometer, a complicate 
data analysis, and high-statistics and low-systematics data collection.    
The latter requirement is rather important since photonuclear   
cross section is small and at its maximum is of only few percents of 
total cross section. 
 
The attenuation method has been initially developed to investigate 
the region of light nuclei where only few final states contribute to 
photonuclear cross section. In that region the photonuclear absorption
is relatively strong  (5-10\% of total cross section). This method 
can also be applied to medium nuclei because of following reasons: \\ 
i) In medium and heavy nuclei the main strength of the GDR is concentrated  
in a narrow energy range. This leads to the relative increase of the  
$\sigma _{nucl}/\sigma _{at}$ ratio in the region of the dipole maximum.\\ 
ii) For these nuclei, the contribution of Compton scattering to atomic
photoabsorption becomes more essential  in respect to  
$e^{+}e^{-}$ pair production. Compton cross section linearly increases with 
the nuclear charge Z. Therefore atomic cross section for these nuclei
raises up slowly than nuclear one which is proportional to $Z^2$.\\ 
 
As it was already mentioned, the main difficulty of the attenuation method is  
the small value of photonuclear cross section. Since the quantity measured
in experiment is total cross section, any  uncertainties
in both experimental data and calculated atomic cross section may affect
final results. The precision of atomic photoabsorption calculations depends on  
used corrections, such as Coulomb and radiative corrections,  
screening of nuclear field  {\it etc.}, and varies for different nuclei  
\cite{at}. Such calculations are more accurate ($\sim$1\%) for light  
nuclei where less corrections are needed  while available experimental 
data provide good testing ground for theoretical approaches. 
The uncertainties raise up for medium and heavy nuclei and reach  2-3\%.   
 
To eliminate such uncertainties from final photonuclear cross section data points,
a phenomenological correction of calculated atomic cross section can be introduced. 
This correction is to be based on obtained total cross section data.
It is known that the nuclear photoabsorption is small 
(about 0.1\% of the total cross section) near ($\gamma,N$) threshold 
and at energies essentially higher the GDR. 
The evaluation of $\sigma_{tot_{nucl}}$ in these regions can be obtained 
from theoretical calculations  and available data on partial channels. 
Further the evaluation of atomic cross section is also possible at the same
energies. The energy dependence of atomic cross section is flat 
and any irregular behavior is not expected. This makes it possible 
to ``re-normalize" the calculated atomic cross section to the measured total 
photoabsorption cross section.  

Another criterion for the correction of atomic cross section is 
the value of integrated photonuclear cross section.
The sum rule for the dominating dipole  E1 nuclear photoabsorption is 

\begin{equation}
\sigma_{E1}^{int} =\int^{\infty}_{0}\frac{d\sigma_{E_{1}}}{dE}dE \approx
 84*\frac{NZ}{A} mb*MeV,
\end{equation} 

\noindent In reality this sum rule is saturated at $\sim 70$ MeV.
Real integrated cross section is slightly larger because of
the contribution of processes others that E1. However the deviation
from (1) is small and can be estimated with reasonable accuracy.

\section{Experiment} 
 
The bremsstrahlung beam, produced in the thin internal tungsten target 
of the C-25 synchrotron of the Institute for Nuclear Research, 
passed a first conical 15 cm thick collimator with the 3 mm diameter
output window, an ionization chamber, a sample of $^{52}Cr$ under study,
a main collimator with a cleaning magnet, and then reached a photon
spectrometer (Fig.1). 
The ionization chamber made it possible the precise (0.2\%) measurement 
of a beam dose. However, the only 
ratio of doses of direct and attenuated spectra was 
needed in this experiment thus reducing the related systematic uncertainty.
Since the number of photons per second in 
the primary beam was very high, the beam intensity was reduced by implanting
a 54.2+60 cm thick aluminum-plus-carbon  absorber. The  
total photoabsorption cross sections in aluminum and carbon have their minimum 
near 20 MeV and strongly raise up below 10 MeV~\cite{at}. Therefore the absorber 
reduced the number of photons at low energies keeping this number at the appropriate
count rate at higher energies.  
 
A $^{52}Cr$ sample was inserted into a computer-controlled trolley.
The trolley could be moved in/out the beam on-line during the data taking. 
In order to reduce systematic uncertainties, the alternate
data collections of the direct and attenuated spectra each lasted 3 and 8 minutes 
respectively, were carried out. The full  measurement  
included hundreds of such cycles resulting in total beam time of about 
500 hours. Thanks to alternate data collection,  any apparatus and accelerator 
instabilities affected  both spectra
in the same way. Consequently the ratio of those spectra 
remained much less affected.

One important problem of such an experiment is    
the detection of secondary particles originating 
from the shower production in a sample, collimators, 
air {\it etc}. Such particles,  after being  re-scattered, 
could reach the detector. To reject this background 
the 50 cm long collimator of 3 mm diameter was installed at 6 m far
from the sample of $^{52}Cr$. This collimator limited the 
detector acceptance to only
of $10^{-6}$ str. The collimator was accurately aligned along the beam axis 
by using a laser. The cleaning magnet located just behind 
the collimator was used to eliminate the remaining $e^+e^-$ pairs.  
  
The photon spectrometer was surrounded by a shield made of concrete blocks, 
polyethylene, cadmium foil and lead. The main component of the spectrometer
was a large 30.5 cm long and 25.4 diameter NAI(TL) crystal manufactured by 
Nuclear Enterprises Ltd. The crystal was viewed by six EMI9758B phototubes.
Signals from the output of the spectrometer were amplified, formed and then
encoded by a charge-to-digit converter (QDC). Measured spectra
were accumulated in a CAMAC buffer memory and recorded on-line.
 
The gains of the phototubes were stabilized by using light pulses from 
thermostabilized light emitting diodes (LEDs). The light pulses were 
generated in time between synchrotron bunches in order to avoid their
overlapping with photons. The recorded spectrum
of light pulses contained two narrow peaks corresponding
to two different amplitudes of current pulses triggering LEDs . 
Calculated mean positions of these peaks were used to correct 
the gain and the pedestal of QDC. This operation 
was activated on-line each 30 seconds. As byproduct, the additional 
illumination of the phototubes has strongly reduced jumping gain 
instabilities related to the difference in count rates (about 7 times) 
in the direct and attenuated spectra. 
In addition, the spectrometer was calibrated each 2-3 hours 
using $^{22}Na$ radioactive source (two $\gamma$-lines of 0.511 and 1.275 MeV).
The resulting short- and long-term instabilities never exceeded 0.2\%. 
 
The duration of a scintillation in the NAI(TL) crystal is 0.8 $\mu$sec. 
If two photons were detected within a time interval shorter than the duration
of a scintillation, such events could be accepted as one event of
larger amplitude. We have developed a special ``pile-up inspector" NIM unit which
analyzed the shape of phototube signals. If the time difference between
two scintillations was less than 0.8 $\mu$sec and more than 0.4 $\mu$sec,
the first signal was recovered and passed to the input of QDC 
while the second was rejected. If the time difference was less than
0.4 $\mu$sec and more than 0.1$\mu$sec, both events were rejected.
Events with time difference shorter than 0.1 $\mu$sec were considered
as one events. The total number of rejected events was recorded. 
Rejected events and remaining pile ups were recovered in an off-line analysis (see below). 
   
A response function of the photon spectrometer and its energy resolution 
were determined by comparing measured and calculated bremsstrahlung 
spectra \cite{af}. A spectrum measured by a photon spectrometer 
is folded with its response function   
 
\begin{equation} 
Sp^{meas}(E)=\int_{0}^{\infty}Sp^{real}(E^{'})A(E^{'},E)dE^{'}  , 
\end{equation} 
 
\noindent $Sp^{meas}$ and $Sp^{real}$ denote measured and real incident spectra, 
$A(E^{'},E)$ is the spectrometer response to a photon of 
energy E. In preliminary calibration tests we have accumulated 60 
bremsstrahlung spectra with different maximum energies varied from 20 to 80 MeV.
The shape of incoming spectra $Sp^{real}$ was calculated taking into account
the photon absorption in the aluminum-carbon
absorber. We have developed an analytical expression describing the response
function of our spectrometer. This expression included 10 parameters each
being a function of energy of incoming photons. These 
parameters were determined from the minimization of $\chi ^2$ between measured 
and calculated spectra. It was found that the spectrometer 
provides the energy resolution of 3.5\% (Full Width at a Half of Maximum) at
photon energies near 20 MeV. At lower and higher energies the resolution 
is slightly worse (5\% and 4\% FWHM at 10 and 40 MeV respectively).

%------------figure 1----------------- 
 
\begin{figure} 
\vspace*{.3cm} 
\centerline{\epsfverbosetrue\epsfxsize=12.5cm\epsfysize=10.5cm 
\epsfbox{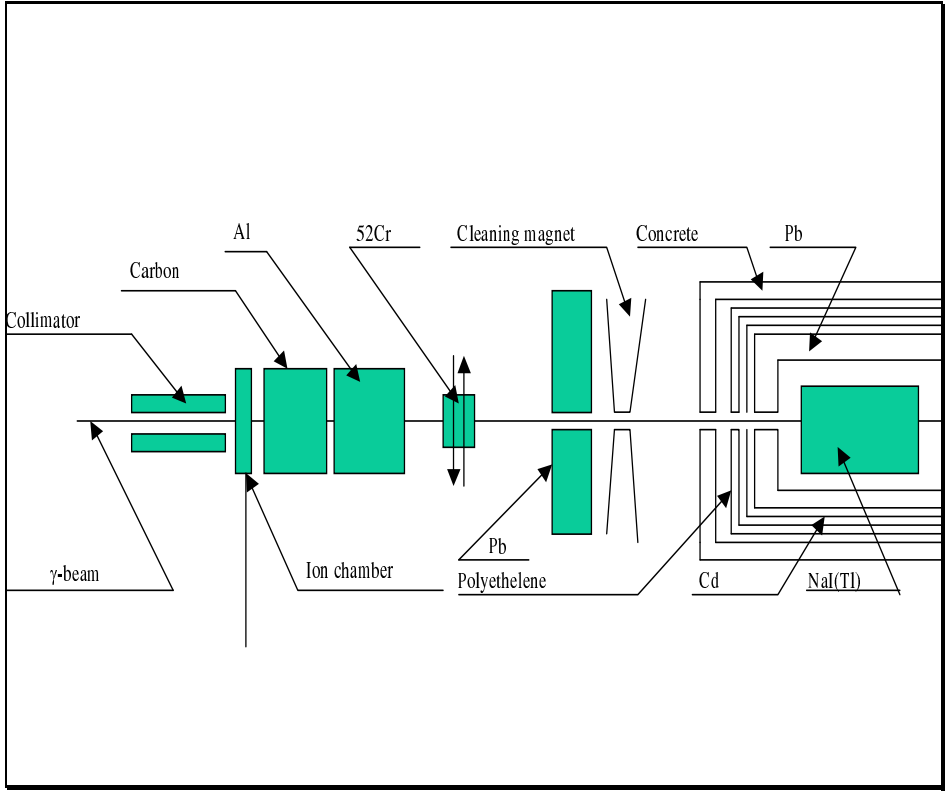}} 
\vspace*{.2cm} 
\caption{Experimental setup.} 
\label{Figure1} 
\end{figure} 
 
%----------end of figure------------- 

----------------------------------------------------------

%-------------figure2---------------- 
 
\begin{figure} 
\vspace*{.3cm} 
\centerline{\epsfverbosetrue\epsfxsize=12.0cm\epsfysize=10.5cm 
\epsfbox{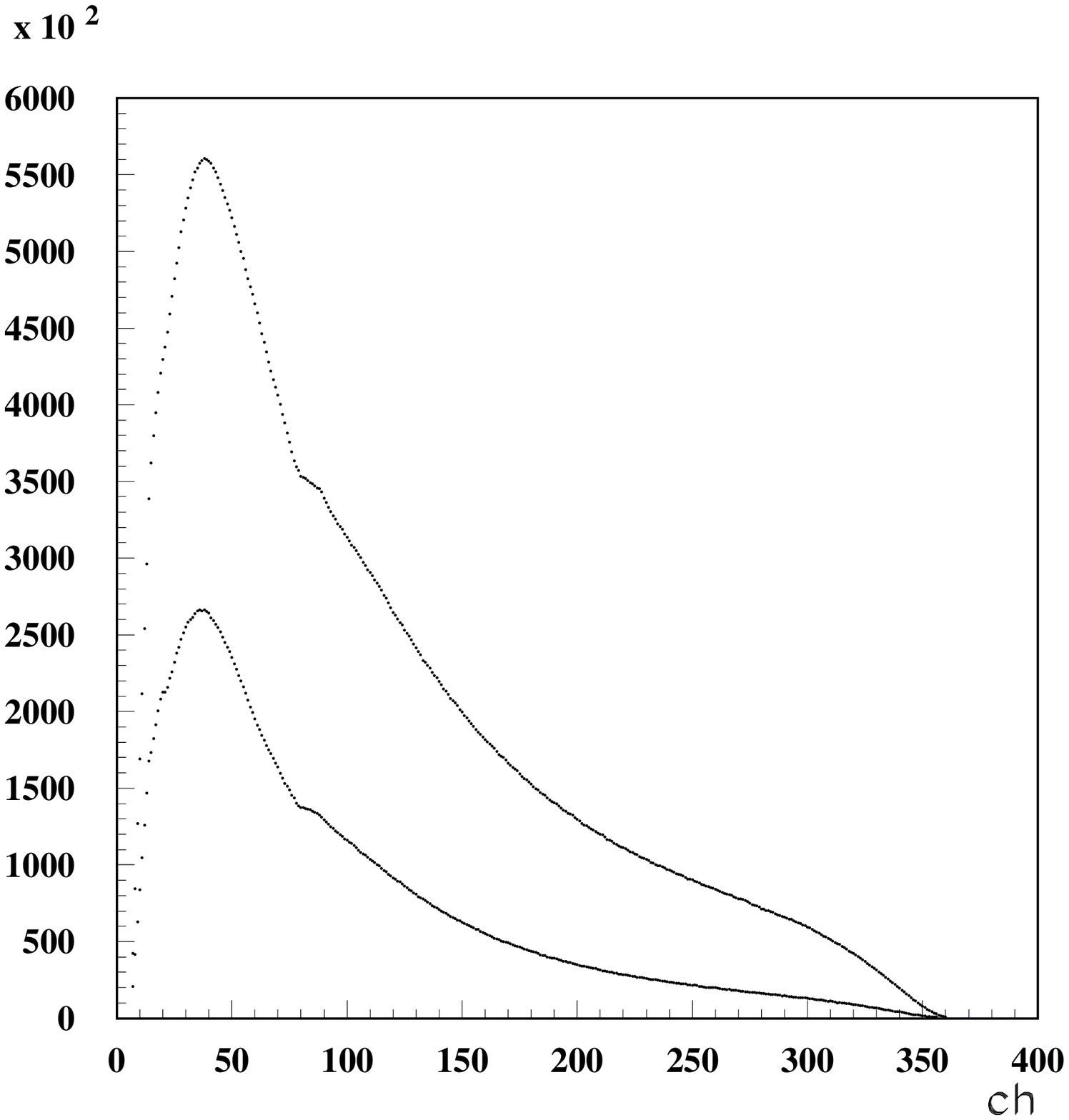}} 
\vspace*{.2cm} 
\caption{Incident photon spectrum (upper curve) and similar spectrum attenuated 
by the sample of $^{52}Cr$ under study. } 
\label{Figure2} 
\end{figure} 
 
%----------end of figure------------- 

\section {Data analysis} 
 
Two measured spectra are shown in Fig.2 These spectra are different
from the spectrum incident on the sample of $^{52}Cr$ and from the attenuated
spectrum because of cosmic background, unrejected pile ups, 
and response function of the spectrometer.  
 
\subsection { Subtraction of cosmic background } 

The average count rate of cosmic particles is much lower than the count rate
of incoming photons. However in the high energy part of the attenuated spectrum 
its contribution  reached 1\% . The spectrum of cosmic background
was measured keeping the accelerator beam off. It was subtracted from
the measured spectra after normalization on the  duration of  
data collection for each spectrum.   
 
\subsection { Pile ups correction} 
 
If two photons were detected within time interval
less than 0.1$\mu$sec, such events were recognized by electronics 
as one event of larger amplitude. 
In the spectrum without attenuation the number of unrejected pile up
events could reach 1-2\% depending on the beam intensity. 
As a result, some false events appeared in the high energy part of 
the spectrum simultaneously with the reduction of its low energy part. 
The correction of the spectra shape
was based on the assumption that the amplitude of a pile-up event is 
equal to the sum of amplitudes of overlapping signals while the 
probability of pile ups is proportional the total count rate. 
The latter was monitored and recorded  during data taking.

\subsection { Correction on the response function }

The measured spectra are folded with the response function the photon 
spectrometer. The main measurement was different from the calibration test:
accelerator bunches were stretched in time in order to decrease the number of
rejected events and pile ups and to reduce significantly the
requested beam time. This mode of operation was achieved by keeping constant
the magnetic field in the accelerator volume after it reaches certain 
level during the acceleration cycle. Once the magnetic field was fixed,
the radius of the electron trajectory inside
the accelerator chamber is defined by the electron kinetic energy and
slowly increases because of acceleration. The passage of the electron beam through
the internal target of the accelerator  slowed down in respect to the  
usual accelerator operation resulting in much longer accelerator bunches.
The photon spectrum emitted in this operation mode was different from the bremsstrahlung
and its shape cannot be calculated as in case of described above calibration
measurements.
   
%-------------figure3---------------- 
 
\begin{figure} 
\vspace*{.3cm} 
\centerline{\epsfverbosetrue\epsfxsize=11.5cm\epsfysize=10.5cm 
\epsfbox{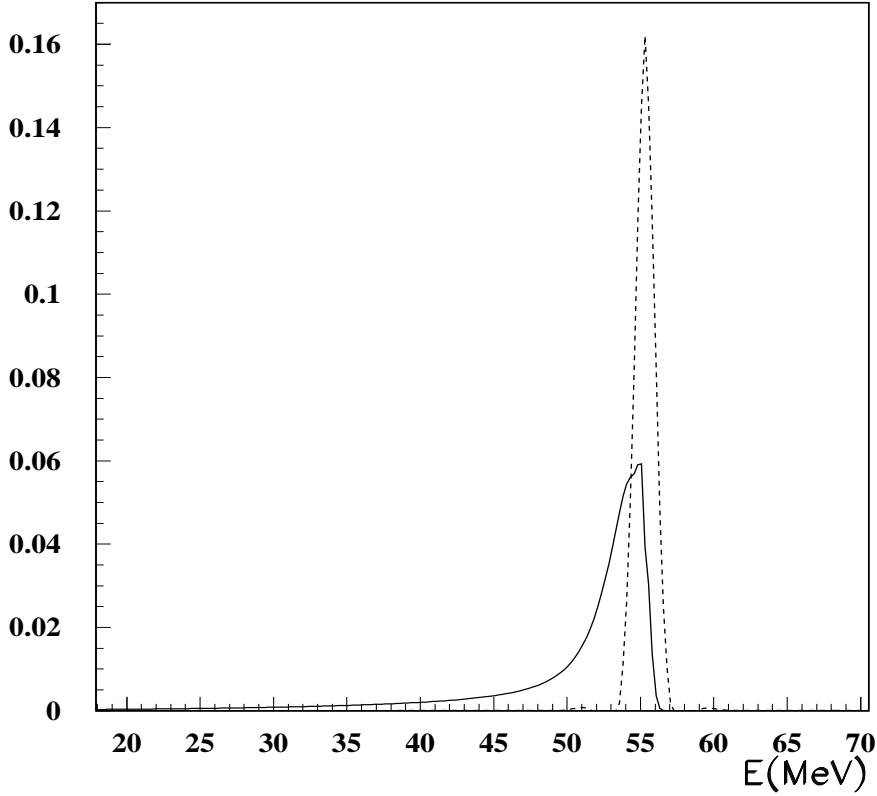}} 
\vspace*{.2cm} 
\caption{ Response function of the NAI(TL) spectrometer 
(solid line) and improved in data analysis  ``new" response function (dashed curve).
Both correspond to 55 MeV photons. } 
\label{Figure3} 
\end{figure} 
 
%----------end of figure------------- 

To recover the distortion of the spectra,
we used the so-called reduction method \cite{rd}. Since in reality
measured spectra are arrays of counts in the QDC channels, 
the expression (2) can be transformed to its discrete form   
 
\begin{equation} 
Sp^{meas}_{i}=\sum_{j=1}^{\infty}Sp^{real}_{j}A_{ij}  , 
\end{equation} 
 
\noindent where indexes i,j denote QDC channels each corresponding to certain
energy, $Sp^{meas}$ denotes a spectrum at the output of the spectrometer, 
$Sp^{real}$ denotes an incident spectrum, $A_{ij}$ denotes the matrix
of the response function. One may define a matrix operator $R_{ki}$ 
which allows to re-calculate $Sp^{meas}$ to another response function  
 
\begin{equation} 
Sp^{new}_{k}=\sum_{i=1}^{\infty}Sp^{meas}_{i}R_{ki} =  
\sum_{j=1}^{\infty}Sp^{real}_{j}\sum_{i=1}^{\infty}A_{ij}R_{ki}  , 
\end{equation} 
 
\noindent In this expression the ``new" spectrum $Sp^{new}$ corresponds to 
the ``new" response function $A^{new}$  
 
\begin{equation} 
A^{new}_{jk} = \sum_{i=1}^{\infty}A_{ij}R{ki}   , 
\end{equation} 
  
\noindent By choosing the operator $R_{ki}$ one may reconstruct the initial
spectrum $Sp^{real}$. In practice, however,
the transformation to an ideal $\delta$-type response function leads 
to the strong increase of statistic errors. It would be  more reasonable to adjust 
the operator $R_{ki}$ such that the resulting response function 
$A^{new}_{ij}$ would be symmetric and would provide 
the energy resolution enough to retrieve an expected structure 
in cross sections. In Fig.3 the real response function for 55 MeV photons and 
the ``new" response function used in data analysis are shown.
Accordingly Fig.4 demonstrates the measured and the reconstructed 
spectra corresponding to the real  and ``new" response functions 
shown in Fig.3.

%-------------figure4---------------- 
 
\begin{figure} 
%\vspace*{.3cm} 
\centerline{\epsfverbosetrue\epsfxsize=11.5cm\epsfysize=10.5cm 
\epsfbox{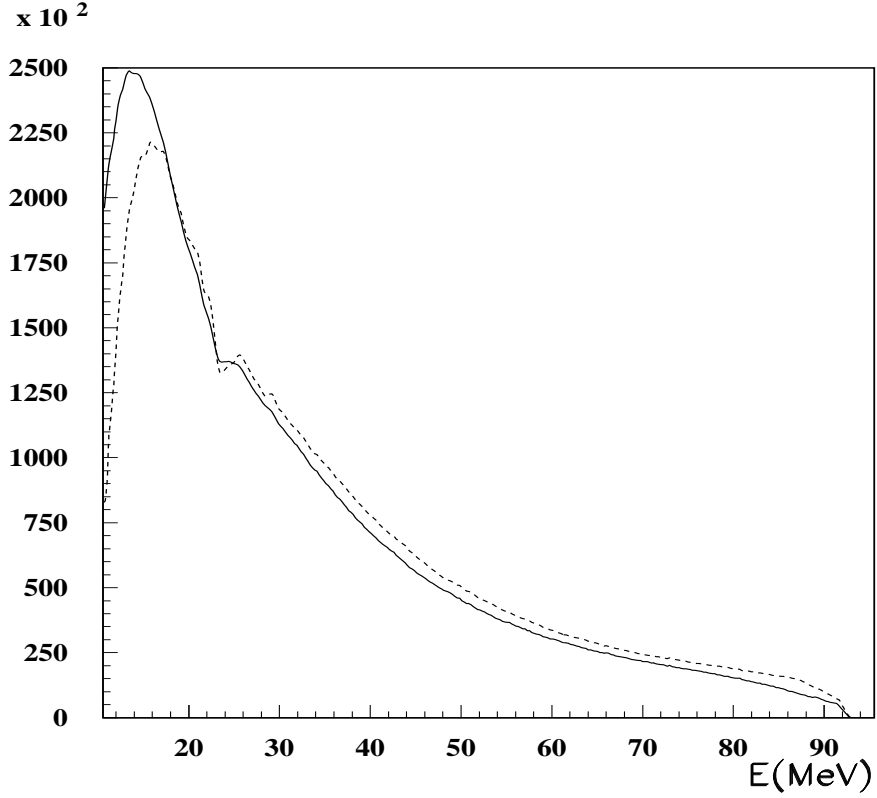}} 
\caption{Measured unattenuated spectrum (solid curve) and same spectrum 
corresponding to the improved ``new" response function.  } 
\label{Figure4} 
\vspace*{.2cm} 
\end{figure} 
 
%----------end of figure------------- 

\section{ Getting the cross section }

A photon flux after passing  a sample of material is attenuated 
in K times

\begin{equation} 
K(E)=\exp{-N_{at}\sigma (E)} , 
\end{equation} 
 
\noindent where $N_{at}$ is the sample length in at/cm$^{2}$, 
$\sigma(E)$ is the total photoabsorption cross section.
$N_{at}$ can be easily calculated from the linear length and the density
of a sample.  In our experiment the attenuation coefficients were
obtained as a ratio of two spectra
 
\begin{equation} 
K(E)=\frac{Sp_{1}(E)}{\alpha Sp_{2}(E)}  . 
\end{equation} 
		  
\noindent where $Sp_{1}$ is the photon beam spectrum without attenuation,
$Sp_{2}$ is the spectrum attenuated by the $^{52}Cr$ sample,  
$\alpha$ is the ratio of beam doses.
The total cross section is derived as
 
\begin{equation} 
\sigma (E)=\frac{1}{N_{at}} \ln{\frac{\alpha Sp_{2}(E)}{Sp_{1}(E)}} . 
\end{equation}

%-------------figure5---------------- 
% 
\begin{figure} 
\vspace*{.3cm} 
\centerline{\epsfverbosetrue\epsfxsize=11.5cm\epsfysize=10.5cm 
\epsfbox{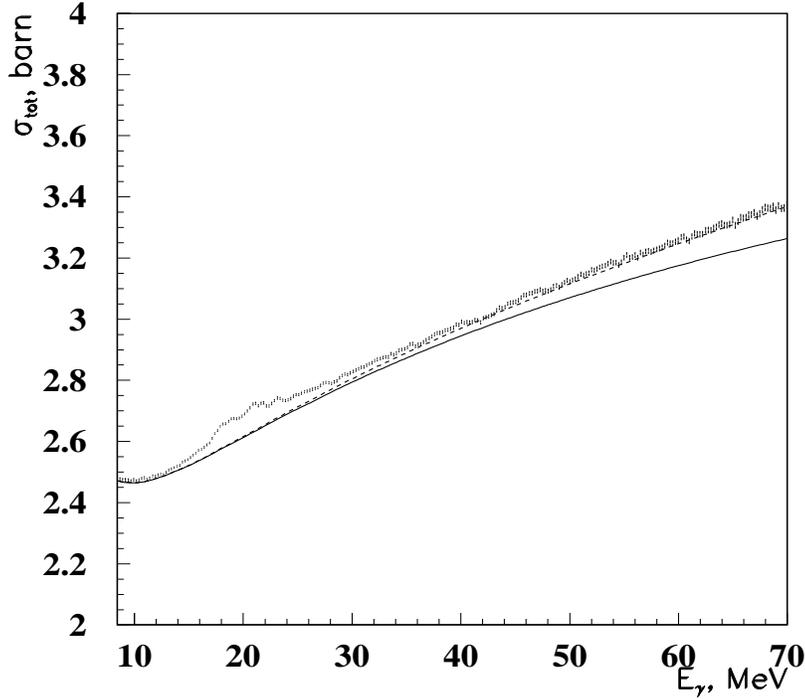}} 
\vspace*{.1cm} 
\caption{ Total photoabsorption cross section of $^{52}Cr$ obtained 
in this experiment. Solid curve shows the atomic cross section
calculated in Ref.~\protect\cite{at}, dashed curve shows the same calculation with
our phenomenological correction. } 
\label{Figure5}
\end{figure} 

% 
%----------end of figure------------- 

The obtained total photoabsorption cross section is shown in Fig.5. 
Data points cover the energy range from 8 to 70 MeV.
These data have been produced with low statistic errors of 0.2\% at energies
of 10-20 MeV and slightly larger, up to 0.6\% at higher energies.
The systematic uncertainty in our data is estimated as 1\%. 
Its main sources are the accuracy of the pile up and 
response function corrections and some of apparatus instabilities.

In Fig.5 the atomic cross section calculated by Hubbell, H.A.Gimm, 
and I.Overbo \cite{at} is compared with our data.  
This model was quite successful in reproducing experimental data for
light ($ A\leq 30$) and heavy ($ A\geq 90$) nuclei.  
If at lower energies there is a good agreement between our data and this model,
at higher energies the discrepancy becomes more pronounced and is out of  
statistic and systematic errors.

%-------------figure6---------------- 
 
\begin{figure} 
%\vspace*{-1cm} 
\centerline{\epsfverbosetrue\epsfxsize=11.5cm\epsfysize=10.5cm 
\epsfbox{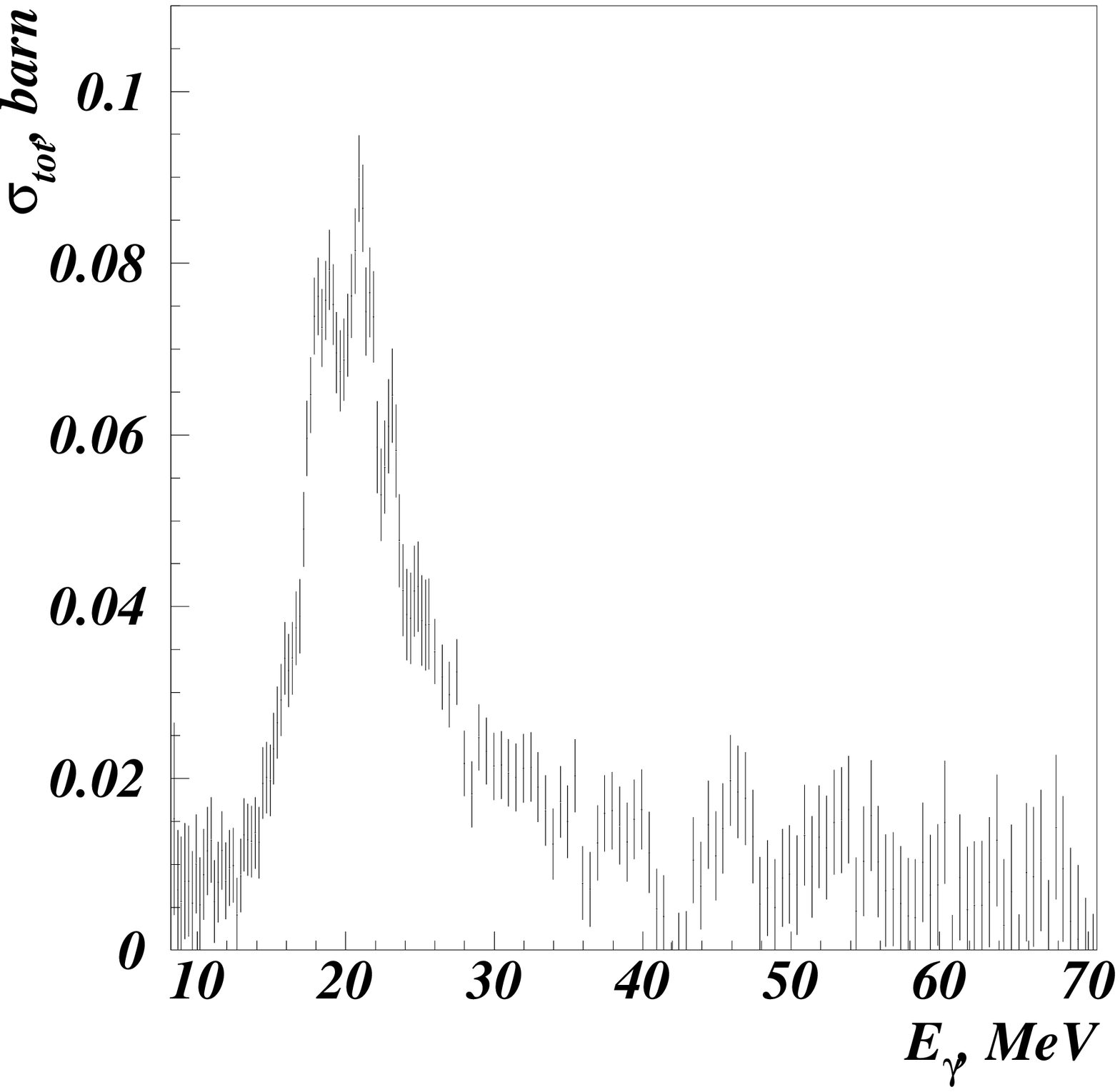}} 
\vspace*{-.2cm} 
\caption{ Total photonuclear photoabsorption cross section of $^{52}Cr$. Solid
curve indicates our calculations which include dipole E1 and quadrupole E2
nuclear photoexcitation, dotted line shows the E2 contribution.} 
\label{Figure6} 
%\vspace*{-0.7cm}
\end{figure} 
 
%----------end of figure------------- 
 
It is well known that nuclear photoabsorption reaches its maximum in the region 
of giant dipole resonances (15-30 MeV) while at lower energies near $(\gamma ,N)$ 
thresholds and essentially higher ($E_{\gamma}\sim$ 60 MeV) its value doesn't 
exceed few mbarn. In order to evaluate the photonuclear cross section, 
we have introduced a small phenomenological correction of the calculated atomic
cross section.

\begin{equation} 
\sigma^{at}_{cor}(E) = \frac{(1+(E-8.4)^{2}}{113000} \sigma^{at}(E) , 
\end{equation} 

\noindent The atomic cross section thus corrected well fits
experimental data (Fig.5). 
 The photonuclear cross section was taken as
a difference between the total and corrected atomic cross sections.

It is worth noting that the introduced correction
is a flat function of energy and cannot affect any structure
in the photonuclear cross section.
In the region of 15-30 MeV, the interaction of photons with nuclei is
governed by the excitation of the E1 giant dipole resonance (Fig.6). 
The dipole bump in the photonuclear cross section 
clearly reveals three peaks at 18.9, 20.9, and 23.1 MeV. At higher energies
there is an indication on a dip-peak structure at 42 - 48  MeV. 
 
\section {Summary and Conclusions} 
 
For the first time the total photoabsorption cross section 
of $^{52}Cr$ nucleus was measured from 8 to 70 MeV of photon energy 
using an attenuation technique. The result deviates from that 
expected on the base of the calculated atomic cross section 
summed with the current estimates for the photonuclear cross section.
This indicates a need to correct the presently available 
calculations of atomic cross sections. Further the reported technique 
to measure total photoabsorption cross sections could be employed
for systematic measurements with various nuclei. Such measurements may 
be useful for many practical applications. 

The derived photonuclear cross section in the region of the giant dipole resonance
exhibits three peaks at 18.9, 20.9, and 23.1 MeV. This correlates with the results
obtained in the study of $\gamma A \to n (A-1)$ reactions in this nuclei domain. 
The peaks might be a manifestation of either the configurational GDR splitting, or 
the isospin splitting, or the combination of both them.
If confirmed, this would be a first observation of the GDR configurational 
splitting in the domain of medium nuclei.
The concise explanation of the observed structure still remains a challenge 
for theoretical interpretation. 

The observed dip-peak structure at 42-48 MEV may signal the photoexcitation of higher-lying 
nuclear resonances. This observation is at the limit of statistical confidence 
and require further experimental confirmation.

It's a pleasure to thank the staff of the C-250 Synchrotron of the Institute 
for Nuclear Research and personally A.M.Gromov for their assistance 
during data taking and the stable beam operation.
This work was supported in part by the Russian Basic Research Foundation 
grant 01-02-16478. 
     
%----------end of main text------------- 
 
%\vspace*{-0.2 cm} 
 
%-------------bibliography---------------- 
 
%----------end of bibliography------------- 
 
\end{document}